\documentclass[headsepline,12pt,a4paper]{amsart}
\usepackage{amsmath,amssymb,amsthm,exscale,calc}
\usepackage{latexsym,textcomp}
\usepackage{parskip}
\usepackage{floatflt}
\usepackage{german}
\theoremstyle{definition}
\newtheorem{Thm}{Theorem}
\newtheorem{Lemma}{Lemma}

\newtheorem{Prop}{Proposition}
\newenvironment{Thm1}{{\noindent\bf Theorem}\;}{}

\newenvironment{Proof}{{\noindent\bf Proof}\;}
{\hfill$\square$\par\medskip} \newlength\headseptemp

\newtheorem*{Remark}{Remark}

\newtheorem{Cor}{Corollary}

\newcommand{\R}{{\mathbb R}}

\newcommand{\N}{{\mathbb N}}

\newcommand{\Lp}{\Delta}
\newcommand{\LC}{\widehat{\Delta}}
\newcommand{\LF}{\widetilde{\Delta}}
\newcommand{\se}{\sigma_{\!\mathrm{ess}}}

\newcommand{\dd}{{\partial}}
\newcommand{\ka}{{\kappa}}
\newcommand{\al}{{\alpha}}
\newcommand{\de}{{\delta}}
\newcommand{\be}{{\beta}}
\newcommand{\gm}{{\gamma}}
\newcommand{\ph}{{\varphi}}
\newcommand{\lm}{{\lambda}}
\newcommand{\eps}{{\epsilon}}
\newcommand{\si}{{\sigma}}
\newcommand{\ra}{{\rightarrow}}

\newcommand{\supp}{{\mathrm{supp}\;}}

\newcommand{\ab}[1]{\left( #1\right)}

\newcommand{\as}[1]{\left\langle #1\right\rangle}

\newcommand{\aV}[1]{\left\Vert #1\right\Vert}

\newcommand{\bs}[1]{\langle #1\rangle}

\newcommand{\bV}[1]{\Vert #1\Vert}

\newcommand{\w}[1]{\widetilde{#1}}

\newcommand{\Hm}[1]{\leavevmode{\marginpar{\tiny%
$\hbox to 0mm{\hspace*{-0.5mm}$\leftarrow$\hss}%
\vcenter{\vrule depth 0.1mm height 0.1mm width \the\marginparwidth}%
\hbox to 0mm{\hss$\rightarrow$\hspace*{-0.5mm}}$\\\relax\raggedright
#1}}}

\begin{document}

\title[The essential spectrum on rapidly branching tessellations]
{The essential spectrum of the Laplacian on rapidly branching tessellations}
\author[M.~Keller]{Matthias Keller}
\address{Fakult\"at f\"ur Mathematik, TU Chemnitz, D -  09107 Chemnitz,
Germany}
\address{currently: Department of Mathematics, Fine Hall
Princeton University Princeton, NJ 08544, USA}

\email{matthias.keller@mathematik.tu-chemnitz.de}
 \maketitle
\def\abstractname{Abstract}
\begin{abstract}
\noindent In this paper we characterize emptiness of the essential
spectrum of the Laplacian under a hyperbolicity assumption for
general graphs. Moreover we present a characterization for emptiness
of the essential spectrum for planar tessellations in terms of
curvature.
\end{abstract}
\addtocounter{section}{-1}
\section{Introduction and main results}
The paper is dedicated to investigate the essential spectrum of the
Laplacian on graphs. More precisely the purpose is threesome.
Firstly we give a comparison theorem for the essential spectra of
the Laplacian $\Lp$ used in the Mathematical Physics community (see
for instance \cite{ASW,AF,AV,Br,CFKS,FHS,GG,Go,Kl,KLPS}) and the
combinatorial Laplacian $\LF$ used in Spectral Geometry (see for
instance
\cite{DKa,DKe,Fu,Wo1}) on general graphs.\\
Secondly we consider graphs which are rapidly branching, i.e. the
vertex degree is growing uniformly as one tends to infinity. We
establish a criterion under which absence of essential spectrum of
the Laplacian $\Lp$ is completely characterized. This criterion will
be positivity of the Cheeger constant at infinity introduced in
\cite{Fu}, based on \cite{Che,D1}. It turns out that in the case of
planar tessellating graphs this positivity will be implied
automatically by uniform growth of vertex degree. Moreover we can
interpret the rapidly branching property as a uniform decrease of
curvature. An immediate consequence is that these operators
have no continuous spectrum.\\
The third purpose is to demonstrate that $\Lp$ and $\LF$ may show a
very different spectral behavior. Therefore we discuss a particular
class of rapidly branching graphs. This discussion will also prove
independence of our assumptions in the results mentioned above. In
the following introduction we will give an overview. We refer to
Section \ref{s:def} for precise definitions.

There is a result of H. Donnelly and P. Li \cite{DL} on negatively
curved manifolds. It shows that the Laplacian $\Lp$ on a rapidly
curving manifold has a compact resolvent, i.e. empty essential
spectrum.

\begin{Thm1}\textbf{(Donnelly, Li)}
\emph{Let $M$ be a complete, simply connected, negatively curved
Riemannian manifold and $K(r) = \sup\{K(x,\pi)\mid d(p,x)\geq r\}$
the sectional curvature for $r\geq0$, where $d$ is the distance
function on the manifold, $p\in M$ and $\pi$ is a two plane in $T_x
M$. If $\,\lim_{r\ra\infty} K(r) =-\infty$, then $\Lp$ on $M$ has no
essential spectrum i.e. $\se(\Lp)=\emptyset$.}\medskip
\end{Thm1}

A remarkable result of K. Fujiwara \cite{Fu} provides an analogue in
the graph case for the combinatorial Laplacian $\LF$.

\begin{Thm1}\textbf{(Fujiwara)}
\emph{Let $G=(V,E)$ be an infinite graph. Then $\se(\LF)=\{1\}$ if
and only if $\al_\infty=1$.}\medskip
\end{Thm1}

Here $\al_\infty$ is a Cheeger constant at infinity. Since the
combinatorial Laplacian $\LF$ is a bounded operator the essential
spectrum can not be empty. Yet it shrinks to one point for
$\al_\infty=1$.

We will show that an analogue result holds for the Laplacian $\Lp$,
which is used in the community of mathematical physicists. Let
$G=(V,E)$ be an infinite graph. For compact $K\subset V$ denote by
$K^c$ its complement $V\setminus K$ and let
$$m_K=\inf\{\deg(v)\mid v\in K^c\}\quad \mathrm{ and}\quad
M_K=\sup\{\deg(v)\mid v\in K^c\},$$ where $\deg:V\ra\N$ is the
vertex degree. Denote
$$m_\infty=\lim_{K\ra\infty} m_K\quad\mathrm{and}\quad M_\infty=\lim_{K\ra\infty} M_K.$$
In the next section we will be precise about what we mean by the
limits. We call a graph \emph{rapidly branching} if
$m_\infty=\infty$. We will prove the following theorems.
\medskip
\begin{Thm}\label{t:Lp_equ}
\emph{Let $G$ be infinite. For all $\lm\in\se(\Lp)$ it holds
$$m_\infty\inf\se(\LF)\leq \lm\leq M_\infty\sup\se(\LF)$$
and
$$\inf\se(\Lp)\leq \min\{m_\infty,M_\infty\inf\se(\LF)\}.$$}
\end{Thm}
In the first statement we have the convention that if
$\inf\se(\LF)=0$ and $m_\infty=\infty$ we set
$m_\infty\inf\se(\LF)=0$. The first part of the theorem shows that
the essential spectra of the operators $\LF$ and $\Lp$ correspond in
terms of the minimal and maximal vertex degree at infinity. The
second part gives two options to estimate the infimum of the
essential spectrum of $\Lp$ from above.

Our second theorem is the characterization of emptiness of the
essential spectrum.
\begin{Thm}\label{t:Lp_DL}
\emph{Let $G=(V,E)$ be infinite and $\al_\infty>0$. Then
 $\se(\Lp)=\emptyset$ if and only if $m_\infty=\infty$.}
\end{Thm}
Note that $m_\infty=\infty$ does not imply $\al_\infty>0$ or
$\se(\Lp)=\emptyset$. This will be discussed in Section
\ref{s:Ex}.\\
We may interpret $\al_\infty>0$ as an assumption on the graph to be
hyperbolic at infinity. (See discussion in \cite{Hi} and the
references \cite{GH,Gr,LS} found there.) Moreover the growth of the
vertex degree can be interpreted as decrease of the curvature. In
this way we may understand Theorem \ref{t:Lp_DL} as an analogue of
Donnelly and Li for $\Lp$ on graphs. For tessellating graphs this
analogy will be even more obvious. Since the continuous spectrum of
an operator is always contained in the essential spectrum there is
an immediate corollary.

\begin{Cor}\label{c:t:Lp_equ}
\emph{Let $G=(V,E)$ be infinite, $\al_\infty>0$ and
$m_\infty=\infty$. Then $\Lp$ has pure point spectrum.}
\end{Cor}

The class of examples for which \cite{Fu} shows absence of essential
spectrum are rapidly branching trees. We will show that the result
is also valid for rapidly branching tessellations. We will formulate
the statement in terms of the curvature because this makes the
analogy to Donnelly and Li more obvious. For this sake we define the
combinatorial curvature function $\ka: V\ra \R$ for a vertex $v\in
V$ as it is found in \cite{BP1,BP2,Hi,Wo2} by
$$\ka(v)=1-\frac{\deg(v)}{2}+\sum_{f\in F,v\in f}\frac{1}{\deg(f)},$$
where $\deg(f)$ denotes the number of vertices contained in a face
$f\in F$. For compact $K\subset V$ let
$$\ka_K=\sup\{\ka(v)\mid v\in K^c\}$$
and $\ka_\infty=\lim_{K\ra\infty}\ka_K$. Obviously
$\ka_\infty=-\infty$ is equivalent to $m_\infty=\infty$. Here is our
main theorem.\medskip

\begin{Thm}\label{tess}
\emph{Let $G$ be a tessellation. Then $\se(\Lp)=\emptyset$ if and
only if  $\ka_\infty=-\infty$. Moreover $\ka_\infty=-\infty$ implies
$\se(\LF)=\{1\}$.}
\end{Thm}
The theorem is a special case of Theorem \ref{t:Lp_DL}. The
hyperbolicity assumption $\al_\infty>0$ follows from
$\ka_\infty=-\infty$ in the case of tessellating graph. In
particular it even holds $\al_\infty=1$ whenever the curvature tends
uniformly to
$-\infty$.\\
Klassert, Lenz, Peyerimhoff, Stollmann \cite{KLPS} proved the
absence of compactly supported eigenfunctions for non-positively
curved tessellations. Since $\ka_\infty=-\infty$ implies
non-positive curvature outside of a certain set $K$ the result
applies here. Hence we have pure point spectrum such that all
eigenfunctions are either supported in $K$ or on an infinite set.

To end this section we introduce a technical proposition which is
used almost throughout all the proofs of the paper. It uses quite
standard technics and may be of independent interest. For a linear
operator $B$ on a space of functions on $V$, we write $B_K$ for its
restriction to the space of functions on $K^c$ with Dirichlet
boundary conditions, where $K\subset V$ is compact set. Let
$l^2(V,g)$ be the space of square summable functions with respect to
the weight function $g$ and $c_c(V)$ the space of compactly
supported functions on $V$.

\begin{Prop}\label{p:B_K}
\emph{Let $G=(V,E)$ be infinite and $B$ a self adjoint operator with
$c_c(V)\subseteq D(B)\subseteq l^2(V,g)$ which is bounded from
below. Then
\begin{eqnarray*}
\inf\se(B)&=&\lim_{K\ra\infty}\hspace{-.3cm}
\inf_{\scriptsize\begin{array}{c}
                   \ph\in c_c(V) \\
                   \supp \ph\subseteq K^c
                 \end{array}}
\hspace{-.3cm}\frac{\bs{B\ph,\ph}_{g}}{\bs{\ph,\ph}_{g}}
\;=\;\lim_{K\ra\infty}\inf\si(B_K), \\
\sup\se(B)&\leq&\lim_{K\ra\infty}\hspace{-.3cm}\sup_{\scriptsize\begin{array}{c}
                   \ph\in c_c(V) \\
                   \supp \ph\subseteq K^c
                 \end{array}}
\hspace{-.3cm}\frac{\bs{B\ph,\ph}_{g}}{\bs{\ph,\ph}_{g}}
\;=\;\lim_{K\ra\infty} \sup\si(B_K).
\end{eqnarray*}
If $B$ is bounded, we have equality in the second formula.}
\end{Prop}

The paper is structured as follows. In Section \ref{s:def} we will
define the versions of the Laplacian which appear in different
contexts of the literature. We discuss Fujiwara's Theorem which can
be understood as a result on compact operators. In Section
\ref{s:Lp} we prove Proposition \ref{p:B_K} and Theorem
\ref{t:Lp_equ} and \ref{t:Lp_DL}. In Section \ref{s:tess} we give an
estimate of the Cheeger constant at infinity for planar
tessellations and prove Theorem \ref{tess}. Finally in Section
\ref{s:Ex} we discuss a class of examples which shows that for
general graphs $\Lp$ and $\LF$ can have a quite different spectral
behavior.

\section{The combinatorial Laplacian $\LF$
in terms of compact operators}\label{s:def}
Let $G=(V,E)$ be a connected graph with finite vertex degree in each
vertex. For a positive weight function $g:V\ra\R_+$ let
\begin{eqnarray*}
   l^2(V,g)&=&\{\ph:V\ra\R\mid \bs{\ph,\ph}_g=\sum_{v\in V}g(v)|\ph(v)|^2<\infty\},\\
     c_c(V)&=&\{\ph:V\ra\R\mid\, |\supp\ph| <\infty\}
\end{eqnarray*}
where $\supp$ is the support of a function. For $g=1$ we write
$l^2(V)$. Notice that $l^2(V,g)$ is the completion of $c_c(V)$ under
$\bs{\cdot,\cdot}_g$.
 For $g=\deg$ it is clear that $l^2(V,\deg) \subseteq
l^2(V)$ and if $\sup_{v\in V}\deg(v)<\infty$ then
$l^2(V)=l^2(V,\deg)$. We occasionally write $l^2(G,g)$ for
$l^2(V,g)$.

For $\ph\in c_c(V)$ and $v\in V$ define the operators
$$(A\ph)(v)=\sum_{u\sim v} \ph(u)\qquad\mathrm{and}\qquad(D\ph)(v)=\deg(v)\ph(v).$$
The operator $A$ is often called the
adjacency matrix. Since we assumed that the graph has no isolated
vertices the operator $D$ has a bounded inverse.

The Laplace operator plays an important role in different areas of
mathematics. Yet there occur different versions of it. To avoid
confusion we want to discuss them briefly. We start with the
Laplacian used in the Mathematical Physicist community in the
context of Schr"odinger operators. For reference see e.g.
\cite{CFKS,D1} (and the references there) or in more recent
publications like \cite{AF,ASW,AV,Br,D2,FHS,GG,Go,Kl,KLPS}.

(1.) The operator $D-A$ defined on $c_c(V)$ yields the following
form
$$\langle d\ph,d\ph\rangle =\frac{1}{2}\sum_{v\in
V}\sum_{u\sim v}|\ph(u)-\ph(v)|^2.$$ The self adjoint operator on
$l^2(V)$ corresponding to this form will be denoted by $\Lp$. It
gives for $\ph\in D(\Lp)$ and $v\in V$
$$(\Lp\ph)(v)=\deg(v)\ph(v) - \sum_{u\sim v}\ph(u).$$
Notice that $\Lp$ is unbounded if there is no bound on the vertex
degree.

We next introduce the combinatorial Laplacian. Two unitary
equivalent versions are found in the literature. They are given as
follows.

(2.) Let $$\widetilde{\Lp}=I-\widetilde{A}=I- D^{-1}A$$ be defined
on $l^2(V,\deg)$, where $I$ is the identity operator. It is easy to
see that $\widetilde{\Lp}$ is bounded and self adjoint. For $\ph \in
l^2(V,\deg)$ and $v\in V$ it gives
$$(\LF\ph)(v)=\ph(v) - \frac{1}{\deg(v)}\sum_{u\sim v}\ph(u).$$
The matrix $\w A$ is often called the transition matrix. This
version of the combinatorial Laplacian can be found for instance in
\cite{DKa,DKe,Fu,Wo1} and many others.

(3.) There is a unitary equivalent version as discussed e.g. in
\cite{Chu}. Let
$$\LC=I-\widehat{A}=I-D^{-\frac{1}{2}}A D^{-\frac{1}{2}}$$ be
defined on $l^2(V)$. It gives for $\ph \in l^2(V)$ and $v\in V$
$$(\widehat{\Lp}\ph)(v)=\ph(v) -
\sum_{u\sim v}\frac{1}{\sqrt{\deg(u)\deg(v)}}\ph(u).$$ Notice that
the operator $$D_{1,\deg}^\frac{1}{2}: l^2(V,\deg)\ra
l^2(V),\quad\ph\mapsto \sqrt{\deg}\cdot\ph, $$ is an isometric
isomorphism and we denote its inverse by
$D_{\deg,1}^{-\frac{1}{2}}.$ Then $$ \LC=
D_{1,\deg}^{\frac{1}{2}}\LF D_{\deg,1}^{-\frac{1}{2}}.$$
Moreover on
$c_c(V)$
$$\Lp=D^{\frac{1}{2}}\LC D^{\frac{1}{2}}.$$

Furthermore we define the Dirichlet restrictions of these operators.
For a set $K\subseteq V$ let $P_K:l^2(V,g)\ra l^2(K^c,g)$ be the
canonical projection and $i_K:l^2(K^c,g)\ra l^2(V,g)$ its dual
operator, which is the continuation by $0$ on $K$. For an operator
$B$ on $l^2(V,g)$ we write
$$B_K=P_K Bi_K.$$
Hence we can speak of $\Lp_K$, $\LF_K$ or $\LC_K$ on $K^c$ with
Dirichlet boundary conditions. Mostly $K$ will be a compact set.
\medskip

For a graph $G$ and compact $K\subseteq V$ define the \emph{Cheeger
constant}, see \cite{Che,DKe,Fu},
\begin{equation*}
\alpha_K=\inf_{W\subseteq K^c,\, |W|<\infty}\frac{|\dd_E W|}{A(W)},
\end{equation*}
where $\dd_E W$ is the set of edges which have one vertex in $W$ and
one outside and $A(W)=\sum_{v\in W}\deg(v)$. Let $W\subseteq K^c$,
for $K$ compact and $\chi$ the characteristic function of $W$. Two
simple calculations, mentioned in \cite{DKa} yield
\begin{equation*}\label{e:cheeger}
\bs{\LF_K\chi,\chi}_{\deg} =\bs{\Lp_K\chi,\chi}=|\dd_E W|
\end{equation*}
and
\begin{equation*}
\bs{\chi,\chi}_{\deg}=\bs{D\chi,\chi}=A(W).
\end{equation*}
This gives
\begin{equation}\label{e:Cheeger}
\al_K=\inf_{W\subseteq K^c,\,
|W|<\infty}\frac{\bs{\LF\chi_W\chi_W}_{\deg}}{\bs{\chi_W,\chi_W}_{\deg}}.
\end{equation}
The set $K(V)$ of compact subsets of $V$ is a net under the partial
order $\subseteq$. We say a function $F:K(V)\ra\R,\;K\mapsto F_K$
converges to $F_\infty\in\R$ if for all $\eps>0$ there is a
$K_\eps\in K(V)$ such that $|F_K-F_\infty|<\eps$ for all $K\supseteq
K_\eps$. We then write $\lim_{K\ra\infty}F_K=F_\infty$. With this
convention we define the \emph{Cheeger constant at infinity} like
\cite{Fu} by
$$\al_\infty=\lim_{K\ra\infty} \alpha_K.$$ The limit always exists
since $\al_K\leq\al_L\leq 1$ for compact $K\subseteq L\subseteq V$.
Therefore we can think of taking the limit over distance balls of an
arbitrary vertex.

The next part is dedicated to a discussion of \cite{Fu}. We will
look at the result from the perspective of compact operators. The
proof is based on two propositions which hold for general graphs. We
present them here as norm estimates on the transition matrix. The
essential part for the proof of the first proposition was remarked
in \cite{DKa}.

\begin{Prop}\label{p:Dodziuk}\emph{For any compact set
$K\subseteq V$ $$\Vert \w A_K\Vert\geq 1- \al_K.$$
In particular $\inf\sigma(\LF_K) \leq \al_K$.}\\
\begin{Proof} By equation (\ref{e:Cheeger}) we
receive $\inf\sigma(\LF_K)\leq \al_K$. Since $\w A_K$ is self
adjoint we get $\inf\sigma(\LF_K)=\inf\sigma(I-\w A_K)=1-\sup\si(\w
A_K)=1- \Vert \w A_K\Vert$ and thus $\Vert \w A_K\Vert\geq 1-
\al_K$.
\end{Proof}
\end{Prop}

\begin{Prop}\label{p:Fujiwara}
\emph{For any compact set $K\subseteq V$ $$\Vert{\w A_K}\Vert\leq
\sqrt{1-\al_K^2}.$$ }
\end{Prop}
The second proposition is derived from the proof of the Theorem in
\cite{DKe}. Alternatively it may be derived from Proposition 1 in
\cite{Fu}. The essential of the proof of this proposition goes back
to Dodziuk, Kendall but the statement can be found explicitly in
Fujiwara. We will use it later, so we state it here as a Theorem.
\begin{Thm}\label{Do}
\emph{For $K\subseteq V$ compact
$$1-\sqrt{1-\al_K^2}\leq \LF_K \leq 1+\sqrt{1-\al_K^2}.$$}
\end{Thm}

\begin{Remark}
(1.) For $K=\emptyset$ we get these estimates for the operator
$\LF$. We can also take the limits over $K$. This is one implication
in Fujiwara's Theorem, since
$\se(\LF)=\se(\LF_K)\subseteq\si(\LF_K)$.
\\
(2.) Since the operators $\LF$ and $\LC$ are unitary equivalent,
similar statements hold for $\LC$ and $\widehat{A}$.
\end{Remark}
The essential parts of the next theorem are already found in
\cite{Fu}. The implications (i.) $\Rightarrow$ (ii.), (iii.)
$\Rightarrow$ (ii.) and (iii.) $\Leftrightarrow$ (iv.) are minor
extensions.

\begin{Thm}\label{LF}
\emph{Let $G$ be infinite. The following are equivalent.
\begin{itemize}
  \item [(i.)] $\se(\LF)$ consists of one point.
  \item [(ii.)] $\se(\LF)=\{1\}$.
  \item [(iii.)] $\w A$ is compact.
  \item [(iv.)] $\lim_{K\ra\infty}\Vert\w A_K\Vert=0$.
  \item [(v.)] $\al_\infty=1$.
\end{itemize}}
\begin{Proof}
The implication (i.) $\Rightarrow$ (ii.) is a consequence of
Proposition \ref{p:B_K} and Theorem \ref{Do}. The implication (ii.)
$\Rightarrow$ (i.) is trivial. Furthermore (ii.) is equivalent to
$\se(\w A)=\{0\}$ which is equivalent to (iii.). Assume (iii.). Let
$(K_n)$ be a growing sequence of compact sets. Choose $f_n\in
l^2(K^c_n,\deg)$, $\aV{f_n}_{\deg}=1$ such that $2\Vert \w A_{K_n}
f_n\Vert_{\deg}\geq \Vert\w A_{K_n}\Vert$. Because $f_n$ is
supported on $K_n^c$ the sequence $(f_n)$ tends weakly to $0$ as
$n\ra\infty$. The compactness of $\w A$ implies $\lim \Vert \w A
i_nf_n\Vert_{\deg}=0$ and thus $\lim\Vert\w A_{K_n}\Vert=0$, which
is (iv.). We assume (iv.), take the limit over $K$ in Proposition
\ref{p:Dodziuk} and conclude (v.). Suppose (v.). For compact $K$ we
have $\w A= i_{K^c}\w A_{K^c}P_{K^c}+ i_K\w A_K P_{K}+C_K$, where
$C_K$ is a compact operator. By Proposition \ref{p:Fujiwara} we have
$\lim\Vert\w A_K\Vert\leq0$. Moreover $\w A_{K^c}$ is compact, since
$K$ is compact. Thus $\w A$ is the norm limit of compact operators
and hence compact, which is (iii.).
\end{Proof}
\end{Thm}

\begin{Remark}
We may think of the problem in an alternative way. For compact
$K\subseteq V$ let $G_K=(V_K,E_K)$ be the graph induced by the
vertex set $K^c$, added by loops in the following way. To each
vertex $v\in K^c$ we add as many loops as there are edges in $\dd_E
K=\dd_E K^c$ which contain $v$. (We say an edge is a loop if its
beginning and end vertex coincides.) We can define projections and
embeddings for $l^2(G,g)$ and $l^2(G_K,g)$ as above. Note that the
projected operators from $l^2(V,g)$ to $l^2(K^c,g)$ and $l^2(G,g)$
to $l^2(G_K,g)$ are unitary equivalent. Thus we can separate the
proof explicitly in graph and operator theory. Proposition
\ref{p:Dodziuk} and \ref{p:Fujiwara} hold for general graphs in
particular also for $G_K$. On the other hand Theorem \ref{LF} is
only operator theory, which uses the estimates on the operator norm
of $\w A_K$.
\end{Remark}

\section{The essential spectrum of $\Lp$}\label{s:Lp}
In this section we compare the operators $\LF$ and $\Lp$. We will
establish bounds on the essential spectrum of $\Lp$ by bounds
obtained for $\LF$. Therefore we will firstly prove Proposition
\ref{p:B_K}. Then we prove two propositions which estimate the
infimum of the essential spectrum of $\Lp$ from below and above.
This will be the ingredients for the proofs of Theorem
\ref{t:Lp_equ} and \ref{t:Lp_DL}.

\begin{Proof}\textbf{of Proposition \ref{p:B_K}}. Without loss of
generality we can assume $B\geq 0$. Let $\lm_0=\inf\se(B)$. Because
$\se(B)=\se(B_K)\subseteq \sigma(B_K)$ it holds $\lm_0\in
\sigma(B_K)$ for any compact $K\subset V$. To show the other
direction we prove that if there is an $\lm \in \si(B_K)\setminus
\se(B)$ then there is $L_0\supset K$ such that $\lm \not\in
\si(B_{L_0})$. It follows $\lm \not\in \si(B_{L})$ for $L\supseteq
L_0$, since $\inf \sigma(B_{L})\geq
\inf\sigma(B_{L_0})$ for $L\supseteq L_0$.\\
For compact $K\subset V$ let $\lm\in \sigma(\Lp_K)$ such that
$\lm<\lm_0$. Choose $\lm_1$ such that
$$\lm<\lm_1<\lm_0.$$
The spectral projection $E_{]-\infty,\lm_1]}$ is a finite rank
operator since $B\geq0$. Let $f_1,\ldots,f_n$ be an orthonormal
basis of the finite dimensional subspace $E_{]-\infty,\lm_1]}
l^2(V,g)$. Now for arbitrary $\eps>0$ choose a compact
$L_\eps\subset V$ so large that for $L\supseteq L_\eps$
$$\max_{j=1,\ldots, n}\bV{P_Lf_j}_g^2
\leq\eps.$$ Let $L\supseteq L_\eps$. For $\ph\in l^2(L^c,g)$ with
$\bV{\ph}_g=1$ there are $\be_1,\ldots,\be_n\in\R$ with
$\be_1^2+\ldots+\be_n^2\leq 1$ such that
$E_{]-\infty,\lm_1],L}\ph=\be_1 P_Lf_1+\ldots+\be_n P_Lf_n$, where
$E_{]-\infty,\lm_1],L}=P_LE_{]-\infty,\lm_1]}i_L$. Remember $P_L$
was the projection of $l^2(V,g)$ onto $l^2(L^c,g)$ and $i_K$ its
dual. Thus
\begin{equation}\label{e:E}
\bV{E_{]-\infty,\lm_1],L}\ph}_g^2 =\be_1^2\bV{P_Lf_1}_g^2+\ldots
+\be_n^2\bV{P_Lf_n}_g^2\leq\eps.
\end{equation}
Now let $\psi\in l^2(L^c,g)$ such that $\bs{B_L\psi,\psi}_g\leq
(\inf\sigma(B_L)+\eps)\bs{\psi,\psi}_g$ and let $d\rho_\psi(\cdot)=
d\bs{B_L E_{]-\infty,\cdot],L}\psi,E_{]-\infty,\cdot],L}\psi}_g$ be
a spectral measure of $B_L$. Then
\begin{eqnarray*}
\bs{B_L\psi,\psi}_g&=&\bs{B_L
E_{]-\infty,\lm_1],L}\psi,E_{]-\infty,\lm_1],L}\psi}_g\\
&&+\bs{B_L E_{]\lm_1,\infty[,L}\psi,E_{]\lm_1,\infty[,L}\psi}_g\\
&\geq& \int_{]\lm_1,\infty[}t \; d\rho_\psi(t)\\
&\geq& \lm_1\int_{]\lm_1,\infty[} 1\; d\rho_\psi(t)\\
&=&\lm_1(\bs{\psi,\psi}_g- \bs{E_{]-\infty,\lm_1],L}\psi,E_{]-\infty,\lm_1],L}\psi}_g)\\
&\geq&\lm_1(1-\eps)\bs{\psi,\psi}_g.
\end{eqnarray*}
In the second step we used that $B$ is positive and in the fifth
step equation (\ref{e:E}). Now we choose $\de>0$ such that
$\lm+\de<\lm_1$. Moreover let
$$\eps=\frac{\lm_1-(\lm+\de)}{\lm_1+1}$$
and $L_0=L_\eps$. By our choice of $\psi$ and $\eps$ we get for all
$L\supseteq L_0$
$$\inf\sigma(B_L)\geq
\frac{\bs{B_L\psi,\psi}_g}{\bs{\psi,\psi}_g}-\eps\geq
\lm_1(1-\eps)-\eps=\lm+\de>\lm.$$ If the operator $B$ is bounded, we
can do a similar estimate from above. Otherwise it still holds
$\sup\se(B)=\sup\se(B_K)\leq\sup\si(B_K)$.
\end{Proof}

Since $\LF$ and $\LC$ are unitary equivalent it makes no difference
to compare to operators $\LF$ and $\Lp$ or the operators $\LC$ and
$\Lp$. Yet $\Lp$ and $\LC$ are defined on the same space, so it
seems to be easier with notation to compare them. However to do this
the following identity is vital. For $\ph\in c_c(K^c)$ one can
calculate
\begin{equation}\label{e:p:Lp_inf}
\frac{\bs{\Lp_K\ph,\ph}}{\bs{\ph,\ph}}=
\frac{\bs{D_K^{\frac{1}{2}}\LC_KD_K^{\frac{1}{2}}\ph,\ph}}{\bs{\ph,\ph}}=
\frac{\bs{\LC_K
D_K^{\frac{1}{2}}\ph,D_K^{\frac{1}{2}}\ph}}{\bs{D_K^{\frac{1}{2}}\ph,D_K^{\frac{1}{2}}\ph}}
\frac{\bs{D_K\ph,\ph}}{\bs{\ph,\ph}}.\\
\end{equation}

\begin{Prop}\label{p:Lp_inf}
\emph{Let $G$ be infinite. Then for $\lm\in \se(\Lp)$
$$ m_\infty\inf\se(\LC)\leq\lm\leq M_\infty\sup\se(\LC).$$}
\begin{Proof} Let $K\subset V$ be compact. By equation (\ref{e:p:Lp_inf})
we have for $\ph\in c_c(K^c)$
$$\frac{\bs{\Lp_K\ph,\ph}}{\bs{\ph,\ph}}\geq
\frac{\bs{\LC_KD_K^{\frac{1}{2}}\ph,D_K^{\frac{1}{2}}\ph}}
{\bs{D_K^{\frac{1}{2}}\ph,D_K^{\frac{1}{2}}\ph}} \inf_{v\in \supp
{\ph}}\!\deg(v).$$ For every $\psi\in c_c(K^c)$ there is an $\ph\in
c_c(K^c)$ such that $\psi =D_K^{\frac{1}{2}}\ph$. Furthermore
$c_c(K^c)$ is dense in the domain of $\Lp_K$ and so we conclude
$$\inf\se(\Lp)=\inf\se(\Lp_K)\geq m_\infty\inf\si(\LC_K).$$
By Proposition \ref{p:B_K} this yields the lower bound. If
$M_\infty=\infty$ the upper bound is infinity. Otherwise by equation
(\ref{e:p:Lp_inf}) $\sup\si(\Lp_K)\leq M_\infty \sup\si(\LC_K)$ and
again by Proposition \ref{p:B_K} we the upper bound.
\end{Proof}
\end{Prop}

\begin{Prop}\label{p:Lp_sup}
\emph{Let $G$ be infinite. Then}
$$\inf\se(\Lp)\leq  \min\{m_\infty,M_\infty\inf\se(\LC)\}.$$
\begin{Proof} Let $v_n\in V$, $n\in\N$ be pairwise distinct such that $\deg
(v_n)\leq m_\infty$. Moreover let $\chi_n$ the characteristic
function of $v_n$. For $K$ compact such that $v_n\in K^c$ it holds
$$\inf_{\ph\in c_c( K^c)}\frac{\bs{\Lp_K\ph,\ph}}{\bs{\ph,\ph}} \leq
\bs{\Lp_K\chi_n,\chi_n}=\deg(v_n)\leq m_\infty.$$ By Proposition
\ref{p:B_K} we have $\inf\se(\Lp)\leq m_\infty.$ On the other hand
we have by equation (\ref{e:p:Lp_inf}) $\inf\si(\Lp_K)\leq
M_K\inf\si(\LC_K).$ By Proposition \ref{p:B_K} we get
$\inf\se(\Lp)\leq M_\infty\inf\se(\LC).$
\end{Proof}
\end{Prop}
\begin{Proof}\textbf{of Theorem \ref{t:Lp_equ}}.
Remember the operators $\LF$ and $\LC$ are unitary equivalent. Thus
$\se(\LF)=\se(\LC)$. The Theorem follows from Proposition
\ref{p:Lp_inf} and \ref{p:Lp_sup}.
\end{Proof}
\begin{Proof}\textbf{of Theorem \ref{t:Lp_DL}}.
By Theorem \ref{Do} we have
$$\inf\si(\LC_K)\geq1-\sqrt{1-\al_K^2}\geq0.$$
Thus by taking the limits Proposition \ref{p:B_K} yields
$\inf\se(\LC)>0$ if $\al_\infty>0$. Propositions \ref{p:Lp_inf} and
\ref{p:Lp_sup} give the desired result.
\end{Proof}
\begin{Remark}
Define $H^1(V)\subseteq l^2(V)$ as the subspace consisting of all
 $f$ with
$$\aV{f}_{H^1}=\aV{f}+\as{df,df}^{\frac{1}{2}}<\infty,$$
where the second term in the sum is the form of $\Lp$ which was
defined in Section \ref{s:def}. Let $j:H^1(V)\ra l^2(V)$ be the
canonical inclusion. Then $\se(\Lp)=\emptyset$ if and only if $j$ is
compact. This can easily be seen by the fact that
$\se(\Lp)=\emptyset$ if and only if $(\Lp^{\frac{1}{2}}+I)^{-1}$ is
compact.
\end{Remark}

\section{Rapidly branching Tessellations}\label{s:tess}

In \cite{Fu} the discussed examples are rapidly branching trees.
Fujiwara showed that for trees $\al_\infty=1$ is implied by
$m_\infty=\infty$. Therefore by Theorem \ref{t:Lp_DL} we have
$\se(\Lp)=\emptyset$ in the case of trees. In this section we want
to extend the class of examples to tessellations. We do this by
showing that $\al_\infty=1$ is implied by $m_\infty=\infty$ for
tessellations as well. For planar graphs tessellations are quite
well understood. We restrict ourselves to the definitions and refer
the reader to \cite{BP1,BP2} and the references contained in there.

Let $G=(V,E)$ be a planar, locally finite graph without loops and
multiple edges, embedded in $\R^2$. We denote the set of closures of
the connected components in $\R^2\setminus \bigcup_{e\in E} e$ by
$F$ and call the elements of $F$ the faces of $G$. We may write
$G=(V,E,F)$ A union of faces is called a \emph{polygon} if it is
homeomorphic to a closed disc in $\R^2$ and its boundary is a closed
path of edges without repeated vertices. The graph $G$ is called a
\emph{tessellation} or \emph{tessellating} if the following
conditions are fulfilled.
\begin{itemize}
  \item [i.)] Any edge is contained in precisely two different faces.
  \item [ii.)] Two faces are either disjoint or intersect in a
  unique edge or vertex.
  \item [iii.)] All faces are polygons.
\end{itemize}
Note that a tessellating graph is always infinite. From now on let
$G=(V,E,F)$ be tessellating. For a set $W\subseteq V$ let
$G_W=(W,E_W,F_W)$ be the induced subgraph, which is the graph with
vertex set $W$ and the edges of $E$ which have two vertices in $W$.
Euler's formula states for a connected finite subgraph $G_W$
\begin{equation}\label{e:Euler}
|W|-|E_W|+|F_W|=2.
\end{equation}
Observe that the $2$ on the right hand side occurs since we also
count the unbounded face. Euler's formula is quite mathematical
folklore, nevertheless a proof can be found for instance in
\cite{Bo}. We denote by $\dd_F W$ the set of faces in $F$ which
contain an edge of $\dd_E W$. In fact each face in $\dd_F W$
contains at least two edges in $\dd_E W$. Therefore $|\dd_F W|\leq
|\dd_E W|$ can be checked easily. Moreover we define for finite
$W\subseteq V$ the \emph{inner degree} of a face $f\in F$ by
$$\deg_W(f)=|f\cap W|$$
Finally let $C(W)$ be the number of connected components in
$G_{V\setminus W}$. Loosely speaking it is the number of holes in
$G_W$.

We need two important formulas which hold for arbitrary finite
subgraphs $G_W=(W,E_W,F_W)$ of $G$. Recall $A(W)=\sum_{v\in
W}\deg(v)$. The first formula can be easily rechecked. It reads
\begin{equation}\label{e:E_W}
A(W)=2|E_W|+|\dd_E W|.
\end{equation}
As for the second formula note that $F_W$ has faces which are not in
$F$. Nevertheless
$$|F_W|-C(W)=|F_W\cap F|.$$
This is the number of bounded faces which are enclosed by edges of
$E_W$. Thus sorting the following sum over vertices according to
faces gives the second formula
\begin{equation}\label{e:F_W}
\sum_{ v\in W}\sum_{f\in F,f\ni
   v}\frac{1}{\deg(f)}
    ={|F_W|}-C(W)+
   \sum_{f\in\dd_F W}
   \frac{\deg_W(f)}{\deg(f)}.
\end{equation}

\begin{Lemma}\label{l:ddEW}
\emph{Let $G=(V,E,F)$ be a tessellating graph. Then for a finite and
connected set $W\subseteq V$
$$|\dd_E W|\geq A(W)-6(|W|+C(W)-2).$$}
\begin{Proof}
By the tessellating property we have
$$\sum_{f\in\dd_F W} {\deg_W(f)}\geq |\dd_E W|.$$
Moreover $\deg(f)\geq3$ for $f\in F$. Combining this with equation
(\ref{e:F_W}) we obtain
\begin{eqnarray*}
{|F_W|}&\leq& \frac{1}{3}\ab{\sum_{ v\in W}\sum_{f\ni
 v}1-\sum_{f\in\dd_F W}
 {\deg_W(f)}}+C(W)\\&\leq& \frac{1}{3}(A(W)-|\dd_EW|)+C(W).
\end{eqnarray*}
By this estimate, Euler's formula (\ref{e:Euler}) and equation
(\ref{e:E_W}) we obtain
\begin{eqnarray*}
2   \leq |W|-\frac{1}{6}(A(W)-|\dd_EW|)+C(W),
\end{eqnarray*}
which yields the Lemma.
\end{Proof}

\end{Lemma}

Now we give an estimate from below of the Cheeger constant at
infinity.\medskip

\begin{Prop}\label{p:CheeTess}
\emph{ Let $G$ be tessellating. Then
$$\al_\infty\geq 1-\lim_{K\ra\infty}\sup_{v\in K^c}\frac{6}{\deg(v)}.$$
}\begin{Proof} We assume w.l.o.g that the compact sets $K$ are
distance balls. To calculate $\al_K$ we can restrict ourselves to
finite sets $W\subset V$, which are connected. Otherwise we find a
connected component $W_0$ of $W$ such that $|\dd_E
W_0|/A(W_0)\leq|\dd_E W|/A(W)$. Moreover we can choose $W$ such that
$C(W)\leq 2$. Otherwise we find a superset $W_1$ of $W$ such that
$|\dd_E W_1|/A(W_1)\leq|\dd_E W|/A(W)$.
\\
Obviously $A(W)\geq |W|\inf_{v\in W}\deg(v)$. By Lemma \ref{l:ddEW}
$$\frac{|\dd_E W|}{A(W)}
\geq\frac{A(W)-6|W|}{A(W)} \geq1-\frac{6}{\inf_{v\in W}\deg(v)}.$$
Hence we have $\al_K\geq1-\sup_{v\in K^c}6/\deg(v)$. We obtain the
result by taking the limit over all compact sets.
\end{Proof}
\end{Prop}
\begin{Remark}
The relation between curvature and the Cheeger constant can be
presented in more detail than we need it for our purpose here. See
therefore \cite{Ke,KP}.
\end{Remark}

\begin{Proof}\textbf{of Theorem \ref{tess}.}
Let $\ka_\infty=-\infty$. This is obviously equivalent to
$m_\infty=\infty$ which implies $\al_\infty=1$ by Proposition
\ref{p:CheeTess}. Thus by Theorem \ref{LF} and Theorem
\ref{t:Lp_equ} we obtain $\se(\LF)=\{1\}$ and $\se(\Lp)=\emptyset$.
On the other hand Proposition \ref{p:Lp_sup} tells us that
$\se(\Lp)=\emptyset$ implies $m_\infty=\infty$ and thus
$\ka_\infty=-\infty$.
\end{Proof}
\begin{Remark}
The implication that $\se(\Lp)=\emptyset$ follows from
$\ka_\infty=-\infty$ can be also obtained on an alternative way.
Higuchi \cite{Hi} and Woess \cite{Wo2} showed independently that
$\al_K>0$  whenever $\ka_K<0$ for $K=\emptyset$. Since
$m_\infty=\infty$ is implied by $\ka_\infty=-\infty$ we can apply
Theorem \ref{t:Lp_equ} immediately.
\end{Remark}

\section{A further class of rapidly branching graphs}\label{s:Ex}
In this section we want to discuss a class of examples which
demonstrates that $\Lp$ and $\LF$ can show very different spectral
phenomena. In particular this examples prove the independence of our
assumptions in Theorem \ref{t:Lp_DL}.

Let $G(n)=(V(n),E(n))$ be the full graph with $n$ vertices. For
$\gm\geq 0$ and $c\geq1$ let $$N_{\gm,c}:\N\ra\N,\quad n\mapsto n[c
n^\gm],$$ where $[x]$ is the the smallest integer bigger than
$x\in\R$. Denote $N_1=1$, $N_2=\max\{[c],2\}$ and for $k\geq 3$
$$N_k=N_{\gm,c}(N_{k-1}).$$
We construct the graph $G_{\gm,c}$ as follows. We start with
connecting the vertex in $G(N_1)$ with each vertex in $G(N_2)$. We
proceed by connecting each vertex in $G({N_k})$ uniquely with
$[cN_k^\gm]$ vertices in $G({N_{k+1}})$ for $k\in\N$. Obviously
$G_{\gm,c}$ is rapidly branching whenever $\gm>0$ or $c>1$, in fact
$N_k\geq 2^{k-1}$. From another point of view $G_{\gm,c}$ is a
'tree' of branching number $[cN_k^\gm]$ in the $k$-th generation,
where we connected the vertices of each generation with one another.
The next theorem shows a scheme of the quite different behavior of
the sets $\se(\Lp)$ and $\se(\LF)$ for the graphs $G_{\gm,c}$.
\medskip

\begin{Thm}\label{t:Gn}
\emph{For $\gm\geq0$ and $c\geq1$  let $G_{\gm,c}$ be as above.
\vspace{-.2cm}$$\begin{array}{clcllcl} \mathrm{\textit{If}} & \gm=0
& \mathrm{\textit{then}} &\al_\infty\!=\!0 ,
& \inf\se(\LF)=0 & \mathrm{\textit{and}} & \inf\se(\Lp)\leq [c].\\
\mathrm{\textit{If}} & \gm\!\in\,]0,1[ & \mathrm{\textit{then}}
&\al_\infty\!=\!0,
& \inf\se(\LF)=0 & \mathrm{\textit{and}} &  \se(\Lp)=\emptyset.\\
\mathrm{\textit{If}} & \gm=1         & \mathrm{\textit{then}}
&\al_\infty\!=\!\frac{c}{1+c}, &
   \inf\se(\LF)\!\in\,]0,1[ & \mathrm{\textit{and}} &  \se(\Lp)=\emptyset. \\
\mathrm{\textit{If}} & \gm>1 & \mathrm{\textit{then}} &
\al_\infty\!=\!1, & \se(\LF)=\{1\} & \mathrm{\textit{and}} &
\se(\Lp)=\emptyset.
  \end{array}
$$}
\end{Thm}
As mentioned above all graphs $G_{\gm,c}$ are rapidly branching if
$\gm>0$ or $c>1$. The theorem shows the independence of our
assumptions and thus optimality of the result. More precisely the
case $\gm=0$ shows that $m_\infty=\infty$ alone does not imply
$\se(\Lp)=\emptyset$. On the other hand the case $\gm\in\,]0,1[$
makes clear that $\se(\Lp)=\emptyset$ does not imply $\al_\infty>0$.
Moreover when $\gm=1$ we see that $m_\infty=\infty$ and
$\al_\infty>0$ does not imply $\al_\infty=1$. The last case is an
example where $\se(\LF)=\{1\}$ and $\se(\Lp)=\emptyset$ like in the
case of trees and tessellations.

For a graph $G$ denote by $B_n$ the set of vertices which have
distance $n\in\N$ or less from a fixed vertex $v_0\in V$. In our
context choose $v_0$ as the unique vertex in $G(N_1)$.

The intuition behind the theorem is as follows. Let
$S_{n,k}=B_{n}\setminus B_{k}$, $n>k$ and $\chi=\chi_{S_{n,k}}$ its
characteristic function. Then one can calculate
$$\frac{\bs{\Lp_{B_k}\chi,\chi}}{\bs{\chi,\chi}}=\frac{|\dd_E
S_{n,k}|}{A(S_{n,k})}\frac{A(S_{n,k})}{|S_{n,k}|}\sim
\frac{c}{N_n^{1-\gm}+c}N_n^{\gm}(N_n^{1-\gm}+c)=cN_n^{\gm}.
$$
The left hand side might be related to $\inf\si(\Lp_{B_k})$.
Moreover the first factor after the equal sign might be related to
$\al_{B_k}$. If this relation holds true we can control the growth
and the decrease of these terms by $\gm$. For instance
$\inf\si(\Lp_{B_k})$ would increase to infinity although $\al_{B_k}$
tends to zero for $\gm<1$.

We denote for a vertex $v\in B_k$
$$\deg_\pm(v)=|\{w\in S_{k\pm1}\mid v\sim w\}|,$$
where we set $S_{k}=B_{k}\setminus B_{k-1}$ for $k\geq2$. To prove
the theorem we will need the following three Lemmata. In \cite{DKa}
the Lemma 1.15 and its subsequent remark gives the following.

\begin{Lemma}\label{l:DKa}
\emph{Let $G$ be a graph. If $(\deg_+(v)-\deg_-(v))/\deg(v)\geq C$
for all $v\in B_n^c$ then $\al_{B_n}\geq C$.}
\end{Lemma}
With the help of this Lemma we will prove the statements for
$\al_\infty$ on the respective graphs.

\begin{Lemma}\label{l:al}
\emph{Let $c\geq1$.
\begin{itemize}
  \item [1.] If $\gm<1$ then $\al_\infty=0$.
  \item [2.] If $\gm=1$ then $\al_\infty=\frac{c}{1+c}$.
  \item [3.] If $\gm>1$ then $\al_\infty=1$.
\end{itemize}}
\begin{Proof}
We get an estimate from above  by calculating
$$\al_{B_{n-1}}\leq\frac{|\dd_E S_{n}|}{A(S_{n})}=\frac{N_n[c
N_n^\gm]+N_n}{N_n(N_n-1)+N_n[c N_n^\gm]+N_n}=\frac{N_n[c
N_n^\gm]+N_n}{N_n^2+N_n[c N_n^\gm]}.$$ To obtain a lower bound for
$\al_{B_{n-1}}$ we use Lemma \ref{l:DKa} and calculate
$$\inf_{v\in B_{n-1}^c}\!\!\frac{\deg_+(v)-\deg_-(v)}{\deg(v)}
=\inf_{k\geq n} \frac{[cN_k^\gm]-1}{1+N_k+[cN_k^\gm]}=
\frac{[cN_n^\gm]-1}{1+N_n+[cN_n^\gm]}.$$ One gets the desired result
by letting $n$ tend to infinity.
\end{Proof}\end{Lemma}

The next lemma is crucial to show absence of essential spectrum for
$\Lp$ when $\gm>0$.\medskip

\begin{Lemma}\label{l:unbounded}
\emph{Let $\gm>0$ and $\ph_k$ functions in $c_c(B_{k-1}^c)$ such
that $\aV{\ph_k}\leq 1$ and $\bs{\Lp_{B_{k-1}}\ph_k,\ph_k}\leq C$
for all $k\in \N$ and some constant $C>0$. Then
$$\lim_{k\ra\infty} \bV{\ph_k}=0.$$}\\
\begin{Proof}
Choose $\ph_k$, $k\in\N$ as assumed. Denote by $\ph^{(i)}_k$ the
restriction of $\ph_k$ to $S_i=B_i\setminus B_{i-1}$ for $i\geq k$
and choose $m>k$ such that $\supp \ph_k\subseteq B_m$. Then an
estimate on the form of $\Lp_{B_{k-1}}$ reads
\begin{eqnarray*}
\bs{\Lp_{B_{k-1}}\ph_k,\ph_k}
&\geq&\sum_{i=k}^m\sum_{v\in S_i}\sum_{w\in S_{i+1},w\sim v}|\ph_k(v)-\ph_k(w)|^2\\
&\geq&\sum_{i=k}^m\Biggl(\sum_{v\in S_i} [cN_i^\gm] \ph_k^2(v)+
\sum_{w\in S_{i+1}}\ph_k^2(w)\\
&&-2\sum_{v\in S_i}\sum_{w\in
S_{i+1},w\sim v} \ph_k(v)\ph_k(w)\Biggr)\\
&\geq&\sum_{i=k}^m\Biggl( [cN_i^\gm]\sum_{v\in S_i}
   \ph_k^2(v)+ \sum_{w\in S_{i+1}}\ph_k^2(w)\\
&&   -2\biggl([cN_i^\gm]\sum_{v\in
S_i}\ph_k^2(v)\biggr)^{\frac{1}{2}}
   \biggl(\sum_{w\in S_{i+1}}\ph_k^2(w)\biggr)^{\frac{1}{2}}\Biggr)\\
&=&\sum_{i=k}^m \ab{[cN_i^\gm]^\frac{1}{2} \bV{\ph_k^{(i)}}-\bV{\ph_k^{(i+1)}}}^2\\
\end{eqnarray*}
In the second step we used that each vertex in $S_i$ is uniquely
adjacent to $[cN_i^\gm]$ vertices in $S_{i+1}$ for $k\leq i\leq m$
and in the third step we used the Cauchy-Schwarz inequality. We
assumed $\bs{\Lp_{B_{k-1}}\ph_k,\ph_k}\leq C$ and in particular this
is true for every term in sum we estimated above. Moreover
$\bV{\ph_k^{(i+1)}}\leq\bV{\ph_k}\leq 1$ for $k\leq i\leq m$ and
thus
$$\bV{\ph_k^{(i)}}\leq\frac{\sqrt{C}+\bV{\ph_k^{(i+1)}}}{cN_i^\frac{\gm}{2}}
\leq\frac{\sqrt{C}+1}{cN_i^\frac{\gm}{2}}.$$ Set $C_0=(\sqrt{
C}+1)/c$. Since the sequence $(N_i^{-\frac{\gm}{2}})$ is summable we
deduce
$$\bV{\ph_k}\leq\sum_{i=k}^m\bV{\ph_k^{(i)}}
\leq C_0\sum_{i=k}^m N_i^{-\frac{\gm}{2}}\leq C_0\sum_{i=k}^\infty
N_i^{-\frac{\gm}{2}} <\infty.$$ We now let $k$ tend to infinity and
conclude $\lim_{k\ra\infty}\bV{\ph_k}=0$.
\end{Proof}
\end{Lemma}

\begin{Proof} \textbf{of Theorem \ref{t:Gn}.}
From Proposition \ref{p:B_K}, \ref{p:Dodziuk} and \ref{p:Fujiwara}
we can deduce
$$1-\sqrt{1-\al_\infty^2}\leq\inf\se(\LF)\leq\al_\infty.$$
Thus by Lemma \ref{l:al} we get the assertion for $\al_\infty$ and
$\inf\se(\LF)$.\\
If $\gm=0$ we get for the characteristic function
$\chi=\chi_{S_{n,k}}$ of ${S_{n,k}}=B_{n}\setminus B_{k}$, $n>k$
$$\frac{\bs{\Lp_{B_{k}}\chi,\chi}}{\bs{\chi,\chi}}=
\frac{[c](N_{k}+N_n)}{\sum_{i=k+1}^n
N_i}=\frac{[c](\frac{N_k}{N_n}+1)}{1+\sum_{i=k+1}^{n-1}
\frac{N_i}{N_n}}.$$ Hence by taking the limit over $n$ we have by
Proposition \ref{p:B_K} that $\inf\se(\Lp)\leq[c]$.\\
Let $\gm>0$ and let $\ph_k$ be functions in $c_c(B_{k+1}^c)$ such
that $\bV{\ph_k}=1$ and
$$\lim_{k\ra\infty}\bs{\Lp_{B_{k+1}}\ph_k,\ph_k}=\inf\se{(\Lp)}.$$
This is possible by Proposition \ref{p:B_K} and a diagonal sequence
argument. As $\bV{\ph_k}=1$ by Lemma \ref{l:unbounded} the term
$\bs{\Lp_{B_k}\ph_k,\ph_k}$ tends to infinity. Thus the essential
spectrum of $\Lp$ is empty.
\end{Proof}

\textbf{Acknowledgments.} I take this chance to express my gratitude
to Daniel Lenz for all the fruitful discussions, helpful suggestions
and his guidance during this work. I also like to thank Norbert
Peyerimhoff for all helpful hints during my visit in Durham. This
work was partially supported by the German Research Council (DFG)
and partially by the German Business Foundation (sdw).

\end{document}